\documentclass[conference,a4paper,twocolumn]{IEEEtran}

\IEEEoverridecommandlockouts

\usepackage{cite}

\ifCLASSINFOpdf
\else
\fi
\usepackage[cmex10]{amsmath}
\usepackage[ruled,vlined,linesnumbered]{algorithm2e}  
\SetAlgoSkip{}

  \usepackage[caption=false,font=footnotesize]{subfig}
\usepackage{stfloats}
\usepackage{url}

\usepackage{graphicx}
\usepackage{amssymb}
\usepackage{hyperref}
\usepackage{xcolor}
\usepackage[T1]{fontenc}
\usepackage[utf8]{inputenc}
\usepackage{balance}
	
	\usepackage{pgfplots}
  \pgfplotsset{compat=newest}
\newlength\figurewidth

\def\RR{\mathbb{R}}

\def\CC{\mathbb{C}}

\def\x{\vect{x}}
\def\u{\vect{u}}
\def\vv{\vect{v}}
\def\y{\vect{y}}
\def\z{\vect{z}}
\def\w{\vect{w}}

\def\g{\vect{g}}

\def\m{\vect{m}}

\def\Mr{M_\mathrm{R}}

\def\Mh{M_\mathrm{H}}
\def\Ml{M_\mathrm{L}}

\def\One{\vect{1}}

\newcommand{\norm}[1]{\|#1\|}
\newcommand{\abs}[1]{\left\vert#1\right\vert}

\newcommand{\vect}[1]{\mathbf{#1}} 

\newcommand{\prox}{\mathrm{prox}}

\newcommand{\sgn}{\mathrm{sgn}}

\newcommand{\dsdrc}{\ensuremath{\Delta\text{SDR}_\text{c}}}
\newcommand{\sdr}{\ensuremath{\mathrm{SDR}}}

\newcommand{\tc}{\ensuremath{\theta_\mathrm{c}}}

\definecolor{matlabCommentGreen}{RGB}{60,118,61}
\newcommand{\comment}[1]{\textcolor{matlabCommentGreen}{{\hspace{1em}\%\,{#1}}}}

\begin{document}



\title{Audio Declipping with (Weighted)\\Analysis Social Sparsity}


\author{\IEEEauthorblockN{
Pavel Záviška, Pavel Rajmic
}
\IEEEauthorblockA{
Brno University of Technology\\
Faculty of Electrical Engineering and Communications, Dept.\ of Telecommunications\\
Technická 12, 616\,00, Brno, Czech Republic\\
Email: pavel.zaviska@vut.cz, pavel.rajmic@vut.cz
}
\thanks{This work was supported by the project 20-29009S of the Czech Science Foundation (GAČR).}
}


%


\maketitle


\begin{abstract}
We develop the analysis (cosparse) variant of the popular audio declipping algorithm
of Siedenburg et al.\ (2014).
Furthermore, we extend both the old and the new variants by the possibility of weighting the time-frequency coefficients.
We examine the audio reconstruction performance of several combinations of weights and shrinkage operators.
The weights are shown to improve the reconstruction quality in some cases;
however, the best scores achieved by the non-weighted methods are not surpassed with the help of weights.
Yet, the analysis Empirical Wiener (EW) shrinkage was able to reach the quality of a computationally more expensive competitor, the Persistent Empirical Wiener (PEW).
Moreover, the proposed analysis variant incorporating PEW slightly outperforms the synthesis counterpart in terms of an auditorily motivated metric.
\end{abstract}



%
\IEEEpeerreviewmaketitle


\section{Introduction}

The paper deals with the problem of recovering signals degraded by hard clipping.
In such a~case, the input signal $\x=[x_1,\ldots,x_N]$ passes through a~system
as follows:
\begin{equation}
	\label{eq:clipping}
	y_n = \left\{
	\begin{aligned}
	&x_n &\text{for} \hspace{1em} &|x_n| < \tc, \\
	&\tc \cdot \sgn(x_n) &\text{for} \hspace{1em}  &|x_n| \geq \tc,
	\end{aligned}
	\right.
\end{equation}
making the output sample $y_n$ saturated if the respective original value $x_n$ exceeds the clipping threshold $\tc$.
This system is nonlinear but acts elementwise.
The related inverse problem is called declipping.

Audio declipping has been studied for a~long time, leading to a~number of proposed methods.
State-of-the-art approaches rely on prior signal assumptions such as
sparsity
\cite{SiedenburgKowalskiDoerfler2014:Audio.declip.social.sparsity,GaultierKiticGribonvalBertin:Declipping2021}
or compressibility by the non-negative matrix factorization (NMF)
\cite{BilenOzerovPerez2015:declipping.via.NMF,Bilen2018:NTF_audio_inverse_problems}.
Other approaches model the signal as an autoregressive process
\cite{Dahimene2008_declipping,Takahashi2013:Hankel_matrix_declipping}.
For a more thorough review, see the survey
\cite{ZaviskaRajmicOzerovRencker2021:Declipping.Survey} and references therein.
Deep learning approaches still appear rarely today~\cite{Tanaka2022:APPLADE.preprint}.

Declipping based on the NMF is today probably the best performing method,
at least as evaluated in the review
\cite{ZaviskaRajmicOzerovRencker2021:Declipping.Survey}.
However, it has a~huge disadvantage which is its extreme computational cost.
The declipping method of Siedenburg et al.\ \cite{SiedenburgKowalskiDoerfler2014:Audio.declip.social.sparsity}
ranks consistently high in various comparisons, sometimes performing even better than the NMF,
while being computationally affordable.
It is based on signal sparsity with respect to the Gabor representation system
\cite{christensen2008}.
In contrast to other methods, it employs neighborhoods of time-frequency coefficients,
leading to time persistent shrinkage
\cite{kowalski2012social},
in effect reducing musical noise reconstruction artifacts.
Nevertheless, the high restoration quality is achieved also due to using
hand-crafted shrinkage operators.

The study \cite{ZaviskaRajmicSchimmel2019:Psychoacoustics.l1.declipping}
was devoted to the inclusion of auditory modeling into the basic audio declipping problem.
Herein, we concluded that involving straightforward time-frequency coefficient weights, quadratically increasing with frequency,
leads to a~significantly better performance in sparsity-based declipping than delicate modeling of psychoacoustic masking curves.
This surprising success was the first source of inspiration for the present paper:
does the coefficient weighting lead to any performance increase also in social-sparsity-based declipping?

Lately, reports have been available showing that the analysis-based (aka cosparse) audio processing methods
perform slightly better than their synthesis (sparse) counterparts
\cite{Kitic2015:Sparsity.cosparsity.declipping,ZaviskaRajmicMokryPrusa2019:SSPADE_ICASSP,ZaviskaRajmicMokry2021:Audio.dequantization.ICASSP}.
In some cases, the analysis approach outperforms the synthesis approach significantly, see for example
\cite{MokryRajmic2019:Reweighted.l1.inpainting}.
The described fact motivated us to develop the analysis social sparsity declipper
and examine its behavior on audio:
does the analysis modeling
improve the declipping performance, compared with the synthesis modeling as presented in
\cite{SiedenburgKowalskiDoerfler2014:Audio.declip.social.sparsity}?

Section \ref{Sec:Synthesis.social.sparsity.declipper}
sums up the core knowledge about social sparsity-based declipping.
Section \ref{Sec:Analysis.social.sparsity.declipper}
then describes the proposed analysis method and presents the respective numerical algorithm.
Section \ref{Sec:Experiments} introduces the reader to the experiments
(both with and without coefficients weighting),
reports the results, and discusses them.


\section{Synthesis Social Sparsity}
\label{Sec:Synthesis.social.sparsity.declipper}
The background and main ingredients of the synthesis social sparsity declipping,
as presented in \cite{SiedenburgKowalskiDoerfler2014:Audio.declip.social.sparsity},
are the goal of this short summary.
The estimate of the original, non-clipped signal is sought by solving
an optimization problem of the form
\begin{multline}
	\label{eq:Social.sparsity.problem.synthesis}
	\min_{\z}
	\left\{
		\frac{1}{2}\norm{\Mr D\z - \Mr\y }_2^2 
		+ \frac{1}{2}\norm{h(\Mh D\z-\Mh\tc\One)}_2^2 + {} \right. \\
		\left. {} + \frac{1}{2}\norm{h(-\Ml D\z-\Ml\tc\One)}_2^2 + \lambda R(\z)
	\right\},
\end{multline}
where $h$ is the hinge function, a~piecewise linear function for which it holds
$h(u)=-\mathrm{ReLU}(-u)$.
Furthermore, 
$\y\in\RR^N$ is the clipped signal,
$\z\in\CC^P$ represents the signal coefficients and
$D\colon\CC^P\to\RR^N$ is the time-frequency \emph{synthesis} operator~\cite{christensen2008}.
Formulation \eqref{eq:Social.sparsity.problem.synthesis}
penalizes solutions inconsistent with the reliable, unclipped samples
(the first term using the mask $\Mr$)
and solutions inconsistent with the clipping model \eqref{eq:clipping} 
(the second and third terms using the masks $\Mh$ and $\Ml$).
The symbol $\One$ represents a~vector of ones.
The last term is the regularizer $R$ with a~balancing parameter $\lambda$;
this function forces sparse or structured-sparse solutions in the time-frequency domain,
which is a~prior suitable for audio signals.

If variations of the $\ell_1$ norm are put in place of $R$,
the algorithm solving \eqref{eq:Social.sparsity.problem.synthesis}
contains simple shrinkages such as the soft thresholding.
Since the quadratic terms in \eqref{eq:Social.sparsity.problem.synthesis}
are Lipschitz-differentiable, the Fast Iterative Shrinkage-Thresholding Algorithm (FISTA) \cite{beck2009fast} can be used as the numerical solver.

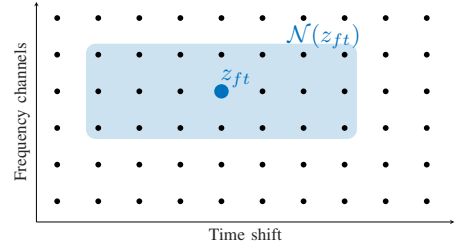
\begin{figure}%
	\centering
	\scalebox{.9}{%
	 
\tikzset{circle node/.style = {circle,inner sep=1pt,draw, fill=white}}
\definecolor{mycolor1}{rgb}{0.00000,0.44700,0.74100}%

 \begin{tikzpicture}[trim axis left, trim axis right, scale=0.6]
	
		\begin{axis}[
		width=4.63in,
		height=2.75in,
		at={(-0.2in,-0.2in)},
		axis lines=left,
		ticks=none,
		xlabel={\large Time shift},
		xlabel style={font=\color{white!15!black}},
		xlabel shift={.1em},
		ylabel={\large Frequency channels},
		ylabel style={font=\color{white!15!black}},
		ylabel shift={.1em},
		]
		\addplot[draw=none] coordinates {(0,0)};
		\end{axis}
	
		\newcommand\xscale{1}
		\newcommand\yscale{0.9}
		
		\fill[rounded corners, color=mycolor1, fill opacity=0.2] (0.7*\xscale, 1.7*\yscale) rectangle (7.3*\xscale, 4.3*\yscale) {};

    \foreach \x in {0,...,9}
    \foreach \y in {0,...,5}
    {
    \fill (\x*\xscale,\y*\yscale) circle (2pt);
    }
		\fill[color=mycolor1](4*\xscale,3*\yscale) circle (5pt);
		\node [color=mycolor1] at (4.35*\xscale, 3.35*\yscale) {$z_{ft}$};
		\node [color=mycolor1] at (6.5*\xscale, 4.5*\yscale) {$\mathcal{N}(z_{ft})$};

  \end{tikzpicture}%
	}
	\vspace{-0.8em}
	\caption{Demonstration of the neighborhood $\mathcal{N}(z_{ft})$ in the TF plane.}%
	\label{fig:neighborhood}%
\end{figure}

The authors of 
\cite{SiedenburgKowalskiDoerfler2014:Audio.declip.social.sparsity}
rely on FISTA but show improved performance when basic shrinkages are
replaced by \emph{empirical} shrinkages, which, by the way,
do not have a counterpart in terms of $R$, as proved in
\cite{Gribonval2018:Characterization.of.prox}.
Specifically, the paper
\cite{SiedenburgKowalskiDoerfler2014:Audio.declip.social.sparsity}
employs four types of shrinkage.
Let $z_{ft}$ represent a~particular time-frequency coefficient.
The shrinkage functions applied to each coefficient in each iteration of FISTA can involve 
LASSO (L), Window Group-LASSO (WGL), Empirical Wiener (EW), or Persistent Empirical Wiener (PEW), such that
\begin{subequations}
	\label{eq:shrinkages}
	\begin{align}
		\text{L}\colon\makebox[.5em]{} & \mathcal{S}_{\lambda,w_{ft}}(z_{ft}) = z_{ft} \cdot \max \left(1-\frac{\lambda\cdot w_{ft}}{\abs{z_{ft}}},0 \right) \label{eq:LASSO.shrinkage} \\
		\text{WGL}\colon\makebox[.5em]{} & \mathcal{S}_{\lambda,w_{ft}}(z_{ft}) = z_{ft} \cdot \max \left(1-\frac{\lambda\cdot w_{ft}}{\norm{\mathcal{N}(z_{ft})}_2},0 \right) \\
		\text{EW}\colon\makebox[.5em]{} & \mathcal{S}_{\lambda,w_{ft}}(z_{ft}) = z_{ft} \cdot \max \left(1-\frac{\lambda^2\cdot w_{ft}}{\abs{z_{ft}}^2},0 \right) \label{eq:EW.shrinkage} \\
		\text{PEW}\colon\makebox[.5em]{} & \mathcal{S}_{\lambda,w_{ft}}(z_{ft}) = z_{ft} \cdot \max \left(1-\frac{\lambda^2\cdot w_{ft}}{\norm{\mathcal{N}(z_{ft})}_2^2},0 \right)  \label{eq:PEW.shrinkage}
	\end{align}
\end{subequations}
where WGL and PEW work with a~coefficient neighborhood~$\mathcal{N}$, as illustrated in Fig.\,\ref{fig:neighborhood}.
EW and PEW are empirical shrinkages.

Above, the weights $w_{ft}$ affect the actual shrinkage thresholds.
For example, \eqref{eq:LASSO.shrinkage} implements soft thresholding with the threshold  $\lambda \cdot w_{ft}$.
There are actually no weights used in
\cite{SiedenburgKowalskiDoerfler2014:Audio.declip.social.sparsity},
which corresponds to the special case of \eqref{eq:shrinkages} with
$w_{ft}=1$ for every pair of $f$ and $t$.
However, we include the weights, because a~part of the experiments utilizes them.

Note that the weights $w_{ft}$ in EW and PEW should correctly come with a~power of two.
In the case of \eqref{eq:EW.shrinkage}, this would make the threshold value effectively $\lambda\cdot w_{ft}$.
In the case of \eqref{eq:PEW.shrinkage}, such a~clear statement cannot be done, since the threshold is modulated by the energy of a~coefficient neighborhood.
While experimenting, we have found that not taking the power in EW and PEW leads to significantly better results.
Such modification can be thus understood as taking the weights $\sqrt{w_{ft}}$ in \eqref{eq:EW.shrinkage}.

In algorithm listings, shrinkages with weights are denoted in a~compact form as $\mathcal{S}_{\lambda,\w}$,
where $\w\in\RR^P,\w>0$.
Alg.\,\ref{alg:social.sparsity} presents the particular numerical steps leading to the solution of 
\eqref{eq:Social.sparsity.problem.synthesis}.

It is also worth noting that any approach stemming from \eqref{eq:Social.sparsity.problem.synthesis}
produces solutions that are generally inconsistent with the clipping model.
Article \cite{ZaviskaRajmicMokry2022:Declipping.crossfading}
shows that a~simple replacement of reliable samples of the solution of
\eqref{eq:Social.sparsity.problem.synthesis}
can significantly improve performance, with negligible costs.

\begin{algorithm}
\SetAlgoVlined
\DontPrintSemicolon
\SetKwInput{Input}{Input}
\SetKwInput{Init}{Initialization}
\SetKwInput{Par}{Parameters}
	\Input{\mbox{$\y,\lambda>0,\tc,\Mr,\Mh,\Ml,D$;
	weights $\w\in\RR^P_+$};
	\newline the shrinkage operator $\mathcal{S}$ (L\,/\,WGL\,/\,EW\,/\,PEW)}
	\Par{$\gamma\in\RR,\ \delta = \norm{DD^*}$}
	\Init{$\hat{\z}^{(0)},\, \z^{(0)} \in \CC^P$}
	%
    \For{$i=0,1,\dots$\,\upshape\textbf{until} convergence}{
			$\g_1^{(i)} = D^* \Mr^* (\Mr D\z^{(i)} - \Mr\y)$ \comment{gradients} \;
			$\g_2^{(i)} = D^* \Mh^*\, h(\Mh D\z^{(i)} - \Mh\tc\One) $ \;
			$\g_3^{(i)} = D^* \Ml^*\, h(-\Ml D\z^{(i)} - \Ml\tc\One) $ \;
			$\g^{(i)} = \g_1^{(i)} + \g_2^{(i)} + \g_3^{(i)} $ \;
			$\hat\z^{(i+1)}\!=\mathcal{S}_{\lambda/\delta,\w} \left( {\z}^{(i)}\! - \frac{1}{\delta} \g^{(i)} \right) $ \hspace{-1.1em}\mbox{\comment{shrinkage}}\hspace{-2em}\;
			${\z}^{(i+1)} = \hat\z^{(i+1)} + \gamma \,( \hat\z^{(i+1)} - \hat\z^{(i)} ) $  \comment{extrapolate} \; 
  }
 	\KwRet{$D\hat{\z}^{(i+1)}$}
	\caption{\mbox{ISTA-type algorithm solving \eqref{eq:Social.sparsity.problem.synthesis}}}
	\label{alg:social.sparsity}
\end{algorithm}


\section{Analysis variant of Social Sparsity}
\label{Sec:Analysis.social.sparsity.declipper}

Among other conclusions,
the audio declipping survey \cite{ZaviskaRajmicOzerovRencker2021:Declipping.Survey}
revealed the fact that analysis (cosparse) variants of reconstruction problems tend to perform slightly better than their synthesis counterparts.
Outstanding results of the synthesis-based social sparsity declipper described
in Sec.\,\ref{Sec:Synthesis.social.sparsity.declipper},
together with its acceptable computation cost,
motivated our effort to develop the analysis variant of the algorithm.

The analysis reformulation of the declipping problem \eqref{eq:Social.sparsity.problem.synthesis} is fairly straightforward and reads
\begin{multline}
	\label{eq:Social.sparsity.problem.analysis}
	\min_{\x}
	\left\{
		\frac{1}{2}\norm{\Mr\x - \Mr\y }_2^2 
		+ \frac{1}{2}\norm{h(\Mh\x-\Mh\tc\One)}_2^2 + {} \right. \\
		\left. {} + \frac{1}{2}\norm{h(-\Ml\x-\Ml\tc\One)}_2^2 + \lambda R(A\x)
	\right\},
\end{multline}
where $\x\in\RR^N$ is the sought time-domain signal
and $A\colon\RR^N \to \CC^P$ is the analysis operator~\cite{christensen2008}.

The composition of the analysis operator with $R$ prevents us from using the FISTA algorithm.
Nevertheless, there are algorithms capable of minimizing optimization problems with composed linear operators
\cite{ChambollePock2011:First-Order.Primal-Dual.Algorithm,Condat2013:PrimalDualSplitting,LorisVerhoeven2011:Generalization.ISTA}.
We stick to the Loris--Verhoeven (LV) algorithm 
\cite{LorisVerhoeven2011:Generalization.ISTA,Combettes2014:FB.of.PD.image.recovery,Condat2019:Proximal.splitting.algorithms}.
The LV algorithm adapted to solving the analysis-based declipping problem 
\eqref{eq:Social.sparsity.problem.analysis}
is proposed in Alg.\,\ref{alg:loris.verhoeven}.

\begin{algorithm}
\SetAlgoVlined
\DontPrintSemicolon
\SetKwInput{Input}{Input}
\SetKwInput{Init}{Initialization}
\SetKwInput{Par}{Parameters}
	\Input{\mbox{$\y,\lambda>0,\tc,\Mr,\Mh,\Ml,A$;
	weights $\w\in\RR^P_+$};
	\newline the shrinkage operator $\mathcal{S}$ (L\,/\,WGL\,/\,EW\,/\,PEW)}
	\Par{$\sigma,\tau\in\RR,\ \rho \in [0; 2-\tau/2]$} 
	\Init{${\x}^{(0)}\in\RR^N, \,\u^{(0)} \in \CC^P$}
	%
    \For{$i=0,1,\dots$\,\upshape\textbf{until} convergence}{
			$\g_1^{(i)} = \Mr^* (\Mr \x^{(i)} - \Mr\y)$ \comment{gradients} \;
			$\g_2^{(i)} = \Mh^*\, h(\Mh \x^{(i)} - \Mh\tc\One) $ \;
			$\g_3^{(i)} = \Ml^*\, h(-\Ml \x^{(i)} - \Ml\tc\One) $ \;
			$\g^{(i)} = \g_1^{(i)} + \g_2^{(i)} + \g_3^{(i)}$ \;
			$\vv^{(i)} = \u^{(i)} + \sigma A(\x^{(i)} - \tau \g^{(i)} - \tau A^*\u^{(i)})$ \comment{aux.} \hspace{-2em} \; 
			$\u^{(i+\frac{1}{2})} = \vv^{(i)} - \sigma\mathcal{S}_{\lambda/\sigma,\w}(\vv^{(i)}/\sigma)$ \;
			$\x^{(i+1)} = \x^{(i)} - \rho\tau(\g^{(i)} + A^*\u^{(i+\frac{1}{2})})$ \;
			$\u^{(i+1)} = \u^{(i)} + \rho^{(i)} ( \u^{(i+\frac{1}{2})} - \u^{(i)} )$
  }
 	\KwRet{$\x^{(i+1)}$}
	\caption{\mbox{Loris--Verhoeven algorithm solving \eqref{eq:Social.sparsity.problem.analysis}}}
	\label{alg:loris.verhoeven}
\end{algorithm}

The shrinkages \eqref{eq:shrinkages} are used in the analysis formulation with no change.
Also the balancing parameter $\lambda$ has the same meaning (and value)
as in Sec.\,\ref{Sec:Synthesis.social.sparsity.declipper}.

Line 7 in Alg.\,\ref{alg:loris.verhoeven} corresponds to the proximal operator
of the Fenchel--Rockafellar conjugate of $\mathcal{S}$,
and we evaluate it indirectly using  the Moreau identity
\cite{BauschkeCombettes2011:Convex.Anal}: 
\begin{equation}
	\prox_{\alpha f^*}(\x) = \x - \alpha\, \prox_{f/\alpha}(\x/\alpha) \ \text{ for  } \alpha \in \RR^+. 
	\label{eq:Moreau_identity}
\end{equation}

The algorithm convergence is guaranteed if $\sigma\tau\|A\|^2 \leq 1$, 
and therefore,
as suggested in \cite{Condat2019:Proximal.splitting.algorithms}, 
we keep $\tau$ as the only tunable parameter and let $\sigma$ be inferred as $1/(\tau\|A\|^2)$.
Since the Lipschitz constant of the gradients of the smooth terms of
\eqref{eq:Social.sparsity.problem.analysis} 
is one,
the step size $\tau$ is limited to $\tau\in(0,2)$.
Another condition is put on the parameter $\rho$; the respective interval is $[0; 2-\tau/2]$.


\section{Experiments and results}
\label{Sec:Experiments}

The experiments were designed to compare the proposed analysis variant of the algorithm with its synthesis counterpart \cite{SiedenburgKowalskiDoerfler2014:Audio.declip.social.sparsity}
in terms of the  quality of restored audio.

The dataset used for the experiments was extracted from the EBU SQAM database,\!\footnote{https://tech.ebu.ch/publications/sqamcd}
which is the same dataset as that used for audio declipping in the survey \cite{ZaviskaRajmicOzerovRencker2021:Declipping.Survey}, as well as in other audio restoration articles 
\cite{ZaviskaRajmicSchimmel2019:Psychoacoustics.l1.declipping,ZaviskaRajmicMokry2021:Audio.dequantization.ICASSP,ZaviskaRajmicMokry2022:Declipping.crossfading}. 
The dataset consists of 10 various music excerpts in mono, sampled at 44.1\,kHz, with an approximate length of 7 seconds.
The clipping threshold \tc{} was determined individually for each audio excerpt based on the input signal-to-distortion ratio (SDR), 
which is  defined for the two signals $\u$ and $\vv$ as
\begin{equation}
\sdr(\u,\vv) = 20\log_{10}\frac{\norm{\u}_2}{\norm{\u-\vv}_2}.
\label{eq:SDR}
\end{equation}
The testing audio excerpts were created by clipping according to \eqref{eq:clipping} using seven clipping levels ranging from 1\,dB to 20\,dB input SDR.

The algorithms were run in MATLAB 2019a using the LTFAT toolbox \cite{LTFAT} for signal synthesis and analysis.
The oversampled STFT with 8,192 samples long Hann window, 75\% overlap and 16,384 frequency channels was used as the signal transform ($A$ and $D$ operators).
The size of the neighborhood $\mathcal{N}(z_{ft})$ was set to 3$\times$7,
(3 coefficients in the frequency direction and 7 in the time direction), 
which seems to be one of the top-performing set-ups \cite{ZaviskaRajmicOzerovRencker2021:Declipping.Survey}.

Both algorithms used an identical setup of common parameters to ensure a fair comparison.
They both exploited the adaptive restart strategy \cite{DonoghueCandes2013:Adaptive.restart}, 
which was proven in \cite{SiedenburgKowalskiDoerfler2014:Audio.declip.social.sparsity} to significantly accelerate the overall convergence.  
It consists in gradually decreasing the value of the balance parameter $\lambda$ every few hundred iterations until the target value of $\lambda$ is reached.
In our experiments, we used 500 inner iterations and 20 outer iterations with $\lambda$ logarithmically decreasing from 10\textsuperscript{$-$1} to 10\textsuperscript{$-$4}.
To slightly accelerate the computations, we also added a~parameter $\varepsilon$, 
which is used to break a~current outer iteration if the $\ell_2$-norm of the difference between the time-domain solution from the current and the previous inner iteration is smaller than $\varepsilon$.
The value of $\varepsilon$ was 0.001.

Other individual parameters of the algorithms are tuned and set such that the respective algorithms deliver the best possible results.
The parameter $\gamma$ of the ISTA-type declipper (Alg.\,\ref{alg:social.sparsity})
changes according to
$(k-1)/(k+5)$,
where $k$ is the inner iteration counter, 
which corresponds to the FISTA acceleration~\cite{beck2009fast}.
The parameters of the Loris--Verhoeven-based declipper (Alg.\,\ref{alg:loris.verhoeven}) were set to $\tau = \text{1.5}, \sigma = \frac{\text{2}}{\text{3}}$, and $\rho = \text{1}$.

\subsection{Comparison of FISTA and LV algorithms}

The quality of the reconstruction is evaluated by two metrics.
The physical similarity between the declipped and the ground-truth waveforms is expressed using the \dsdrc{}, 
which represents the SDR improvement over the degraded signal,
computed only on clipped samples of the waveforms.
Nevertheless, a~similarity of the waveforms may not necessarily imply perceptual quality. 
Therefore, the restoration quality is also evaluated using the PEMO-Q
\cite{Huber:2006a}, which is a perceptually motivated metric for audio quality evaluation. 
The output of PEMO-Q is a~number on the objective difference grade (ODG) scale ranging from \textminus4 (worst quality) to 0 (best quality).
It is worth noting that PEMO-Q has originally been developed for rating audio quality degraded by a~compression, the same as PEAQ
Perceptual Evaluation of Audio Quality (PEAQ) \cite{Kabal2002:PEAQ}.
However, there is no auditorily motivated objective tool available specifically for declipping, and PEMO-Q seems to produce more reliable grades than PEAQ does for declipped audio.
Therefore we stick to PEMO-Q in line with the recent literature.

\begin{figure}%
	\centering
	\subfloat[][L]{\hspace*{-2mm}\includegraphics[width=0.499\columnwidth]{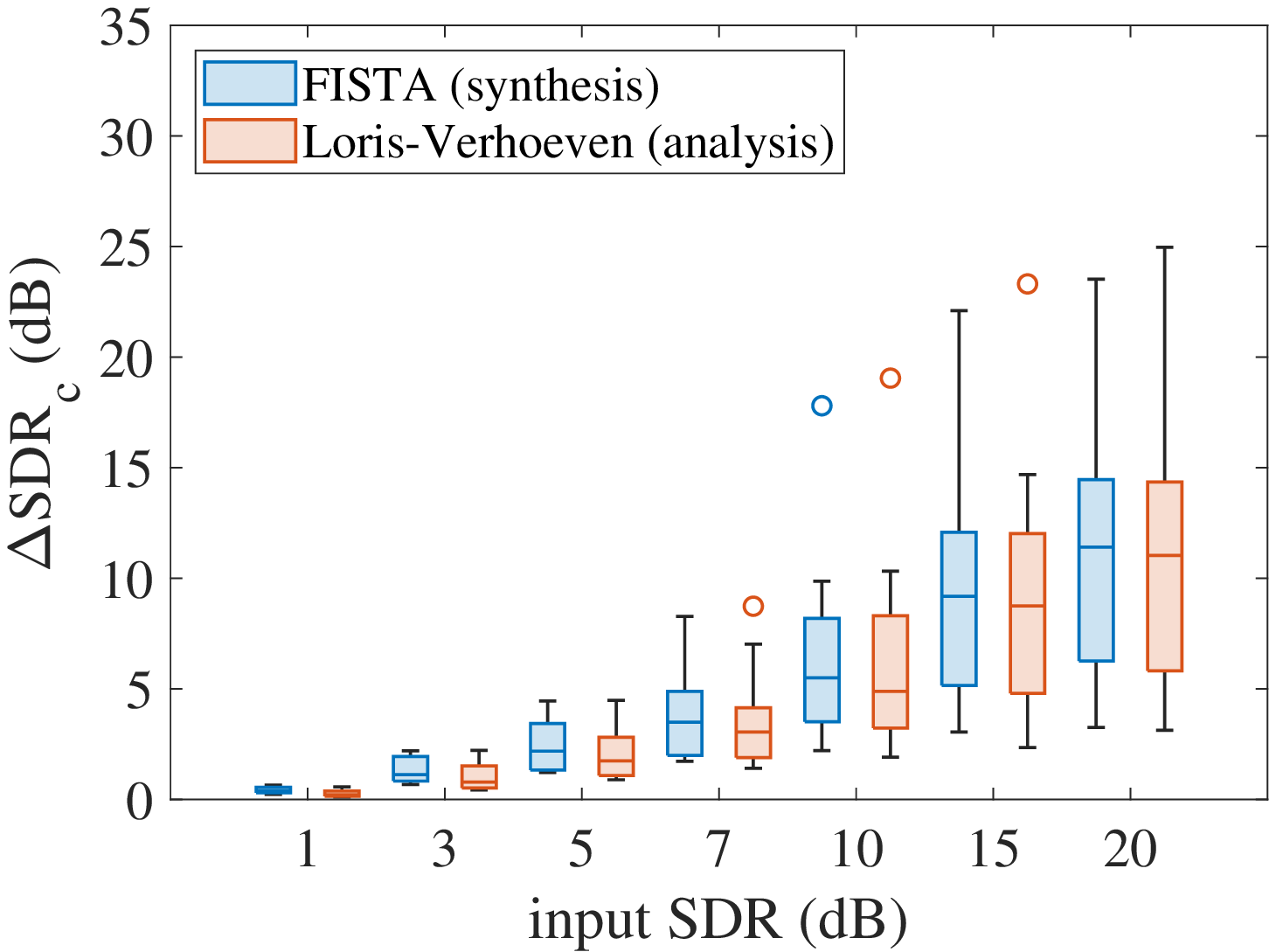}\label{subfig:dSDR_L}}\hfill
	\subfloat[][EW]{\includegraphics[width=0.499\columnwidth]{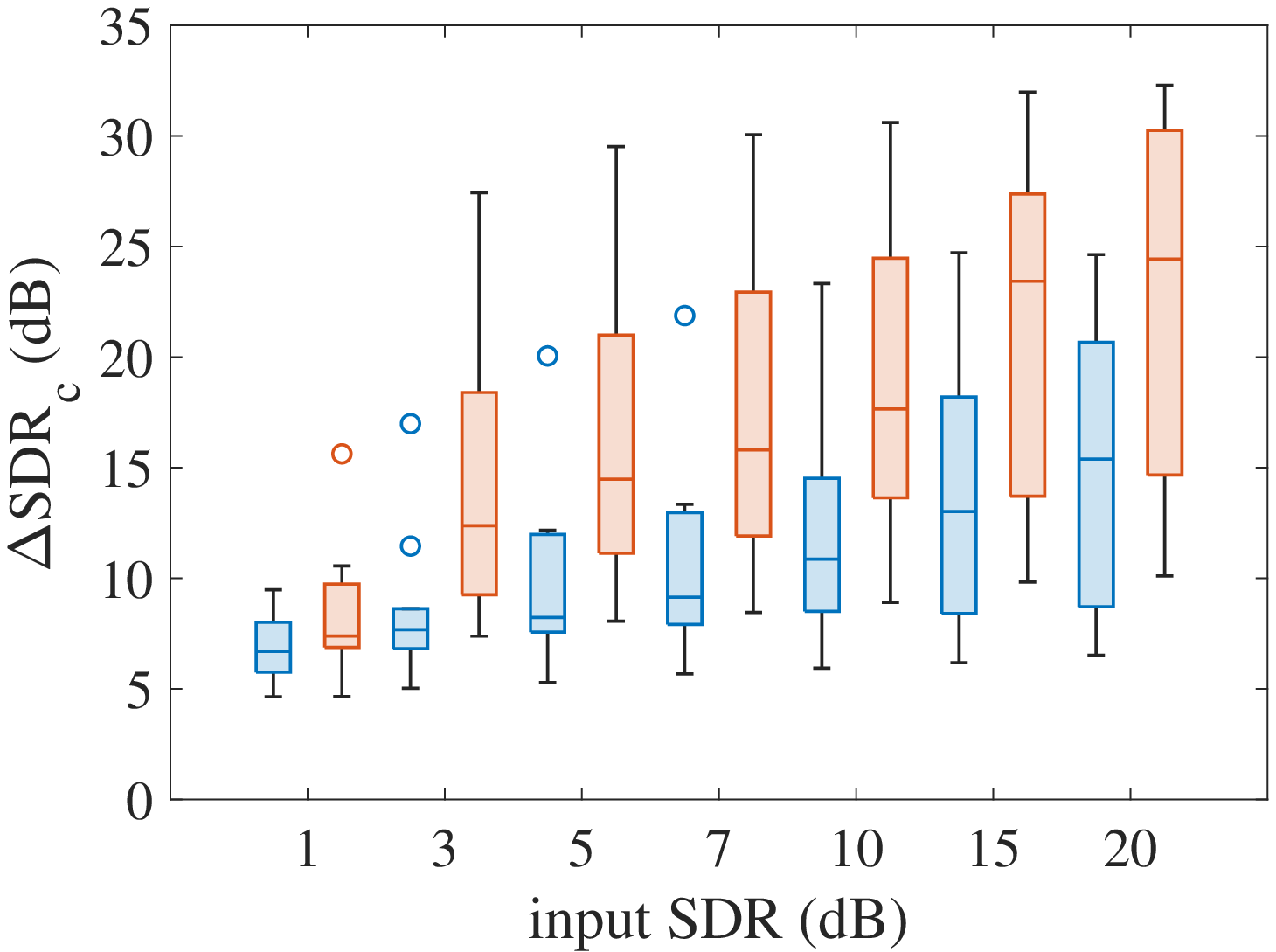}\label{subfig:dSDR_EW}}\\[-0.1cm]
	\subfloat[][WGL]{\hspace*{-2mm}\includegraphics[width=0.499\columnwidth]{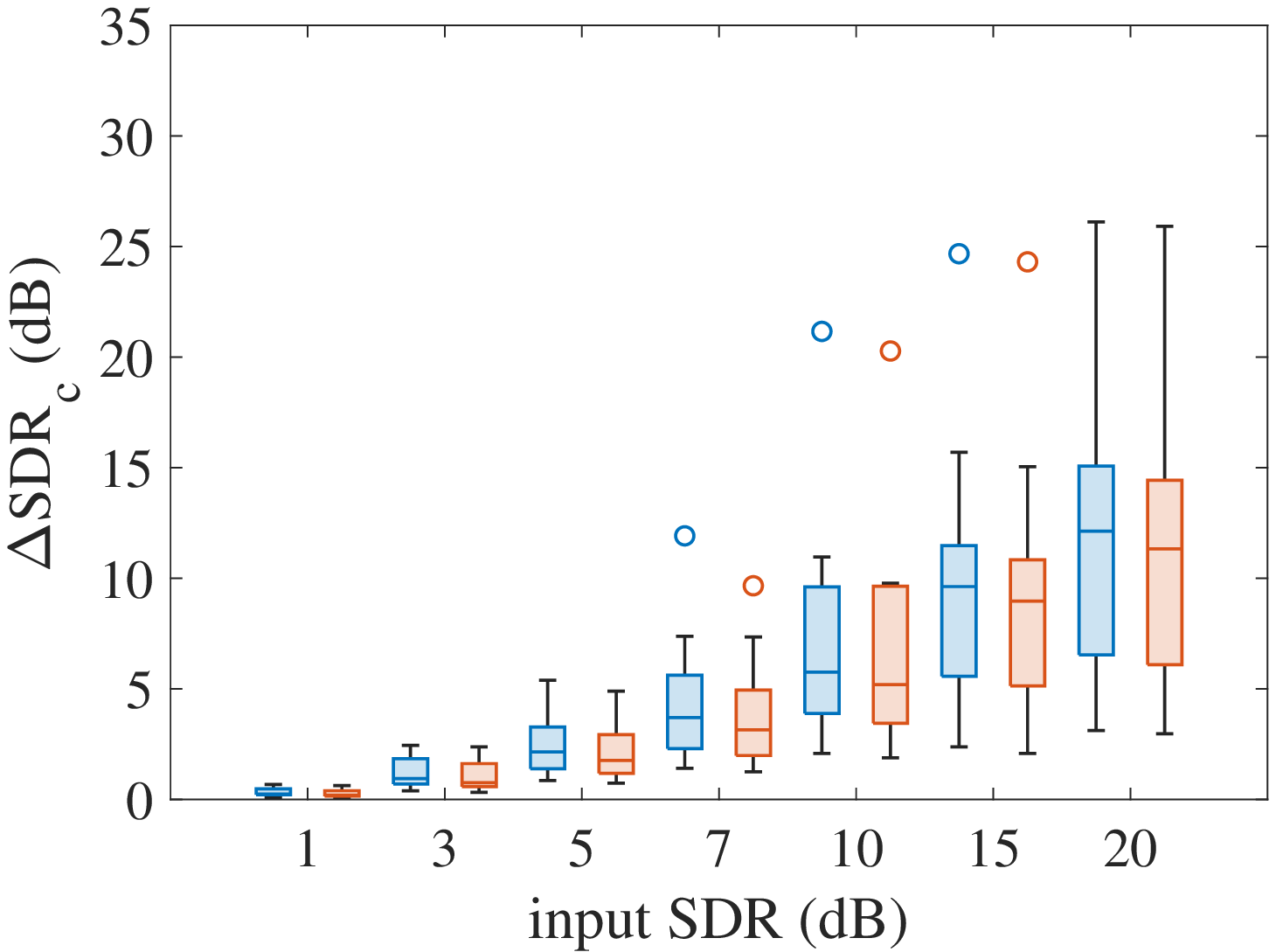}\label{subfig:dSDR_WGL}}\hfill
	\subfloat[][PEW]{\includegraphics[width=0.499\columnwidth]{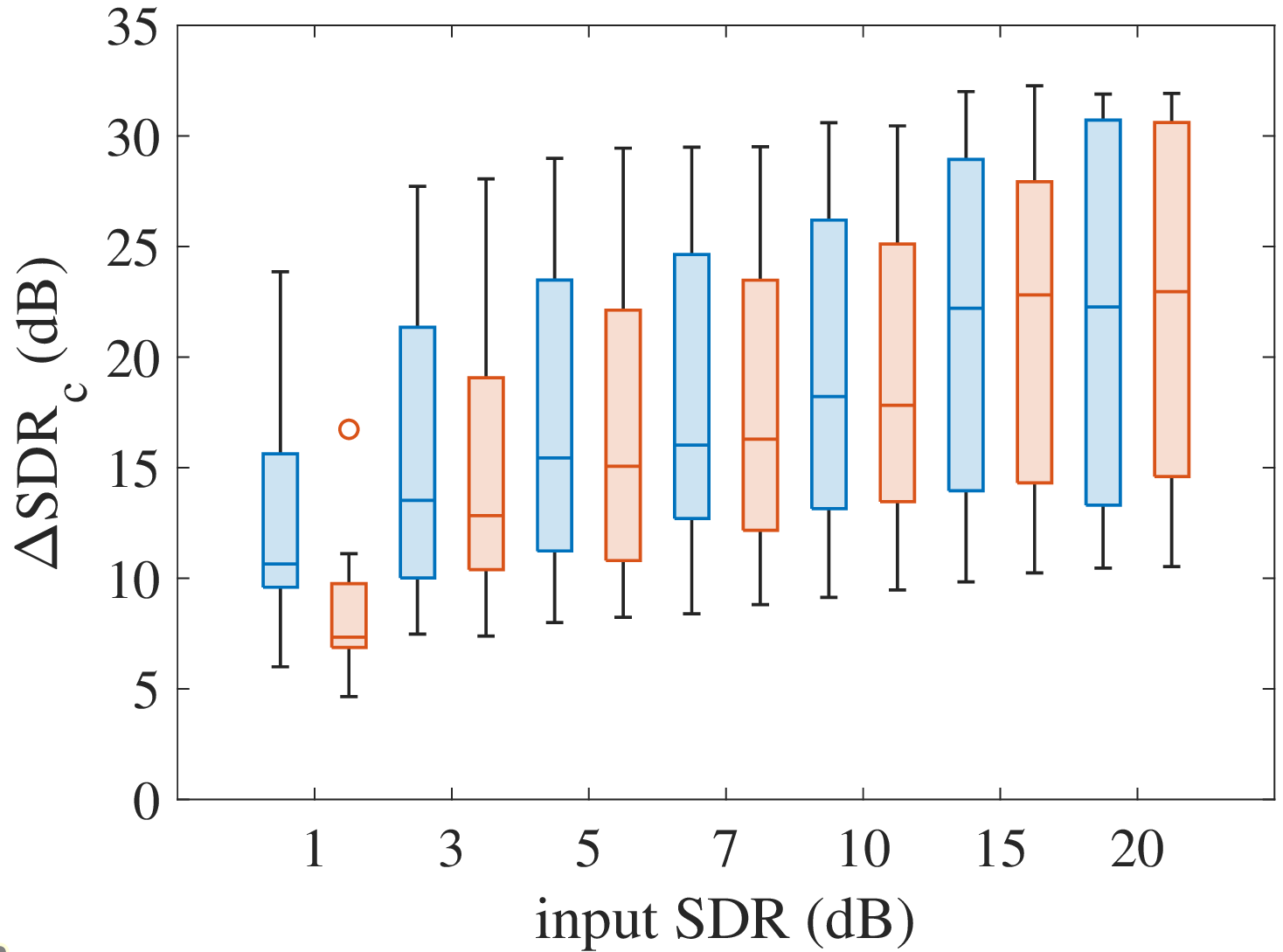}\label{subfig:dSDR_PEW}}
	\caption{Comparison of the synthesis and the analysis approaches to audio declipping for four different shrinkage operators using \dsdrc{}.}
	\vspace{-1em}%
	\label{fig:dSDR_boxplots}%
\end{figure}

The \dsdrc{} values achieved are depicted in Fig.\,\ref{fig:dSDR_boxplots} in the form of box charts,
for the shrinkage operators presented in Equations~\eqref{eq:shrinkages}.
The results show that the proposed analysis approach is slightly worse in the case of L (see Fig.\,\ref{subfig:dSDR_L}) and WGL (see Fig.\,\ref{subfig:dSDR_WGL}) shrinkage types.
However, in the case of EW, it is better by almost 6\,dB on average.
With respect to the PEW shrinkage operator, the analysis variant performs slightly better for higher input SDR values but worse for lower input SDR values, especially for 1\,dB.

\begin{figure}%
	\centering
	\subfloat[][L]{\hspace*{-2mm}\includegraphics[width=0.499\columnwidth]{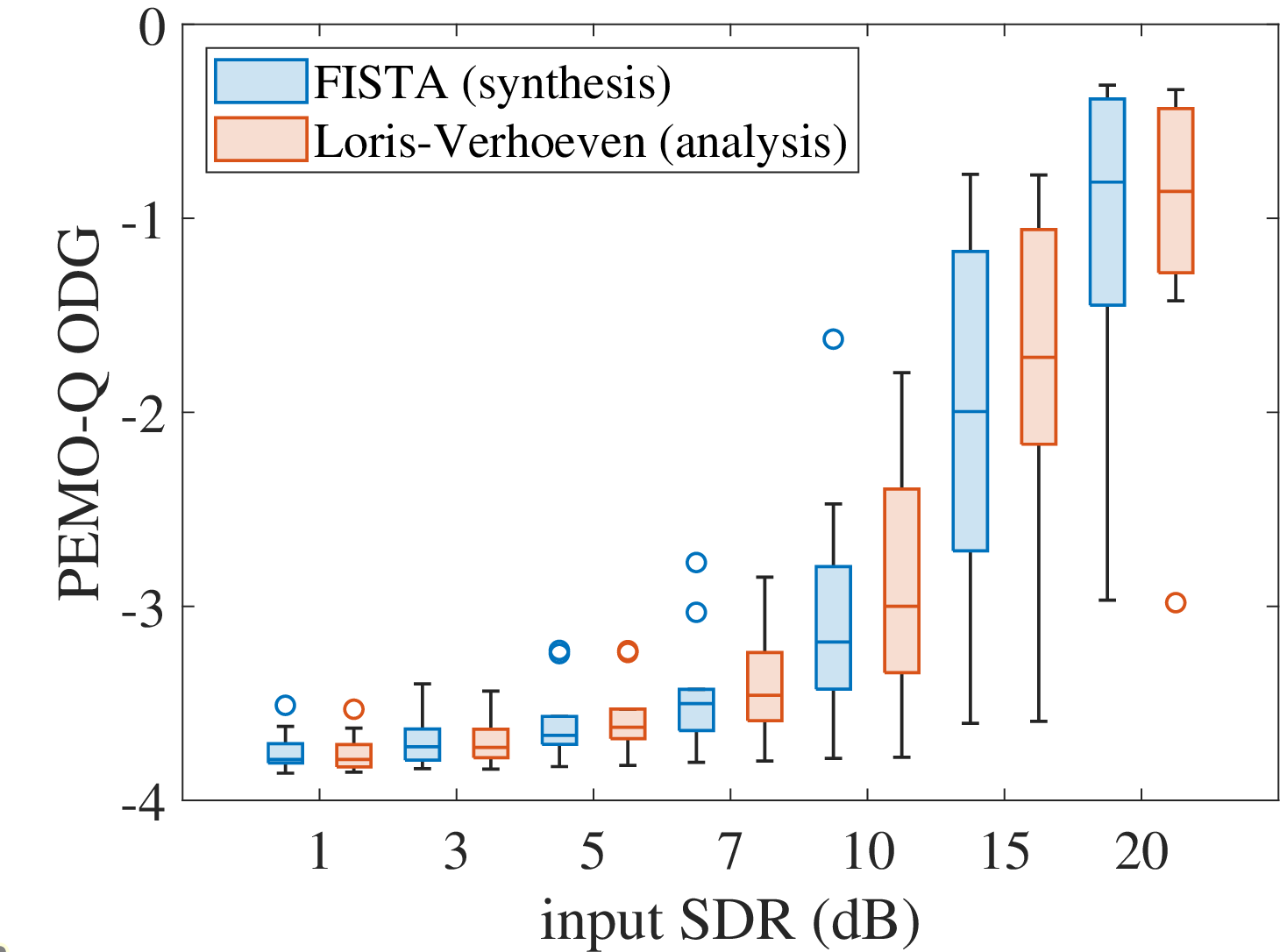}\label{subfig:PEMO_Q_L}}\hfill
	\subfloat[][EW]{\includegraphics[width=0.499\columnwidth]{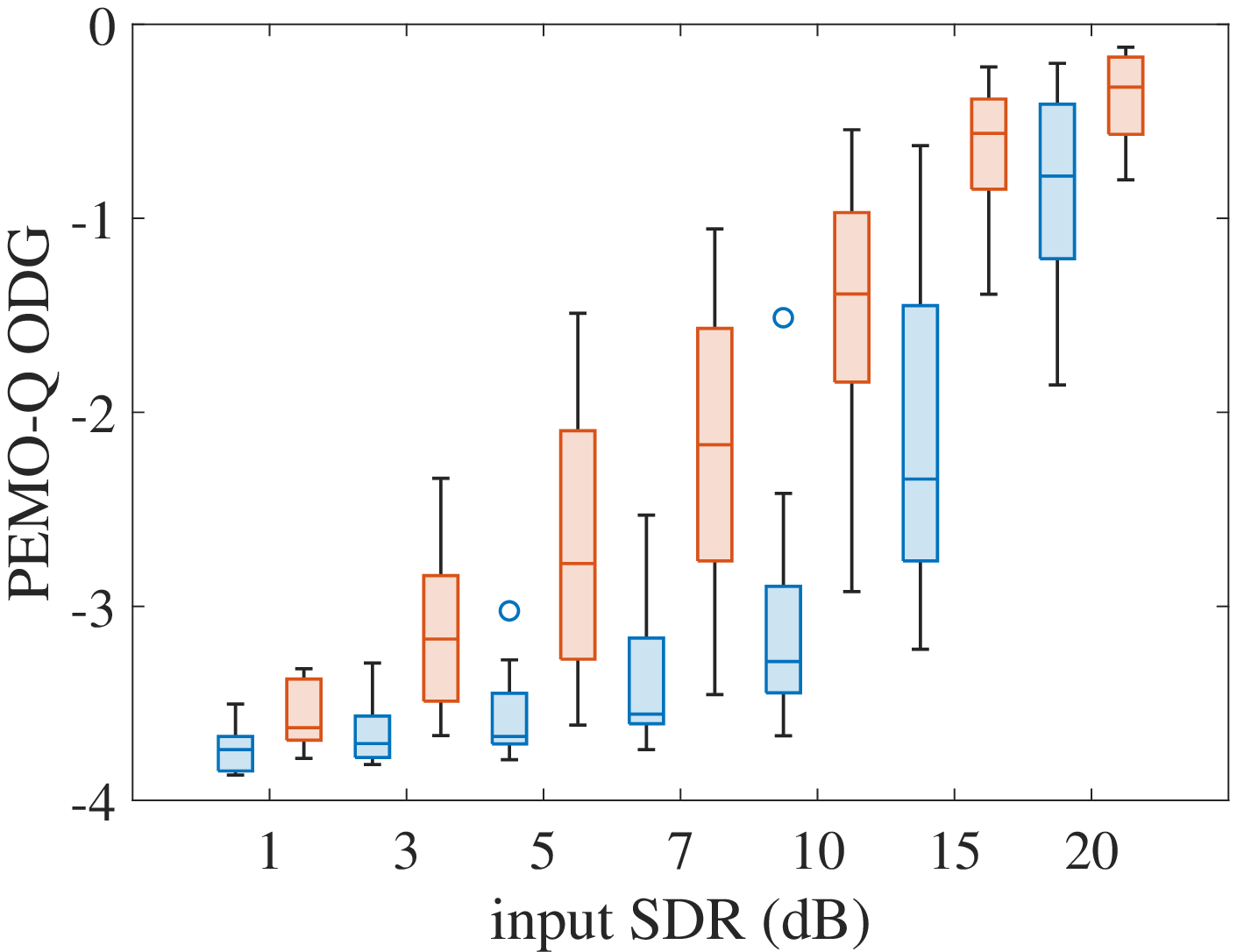}\label{subfig:PEMO_Q_EW}}\\[-0.1cm]
	\subfloat[][WGL]{\hspace*{-2mm}\includegraphics[width=0.499\columnwidth]{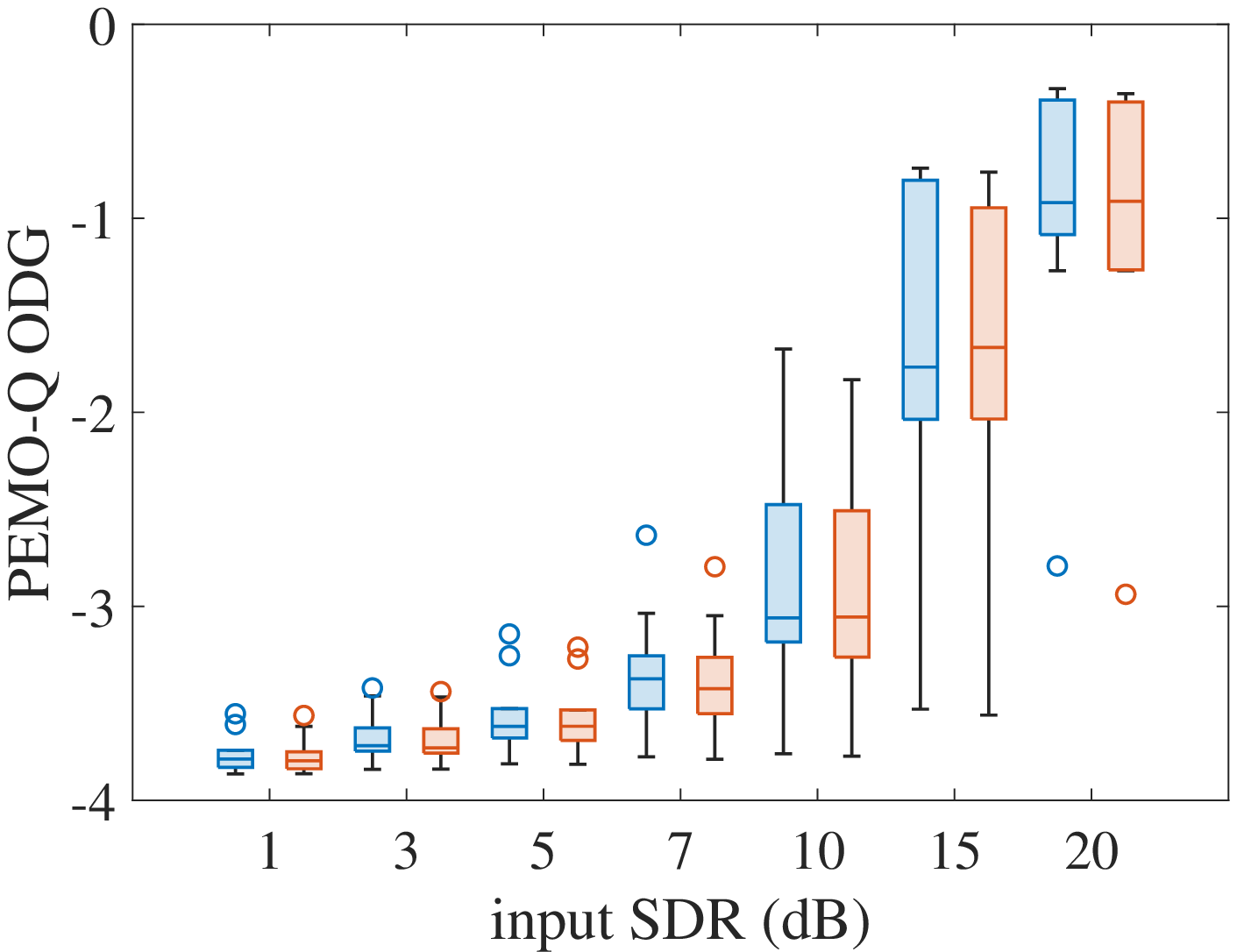}\label{subfig:PEMO_Q_WGL}}\hfill
	\subfloat[][PEW]{\includegraphics[width=0.499\columnwidth]{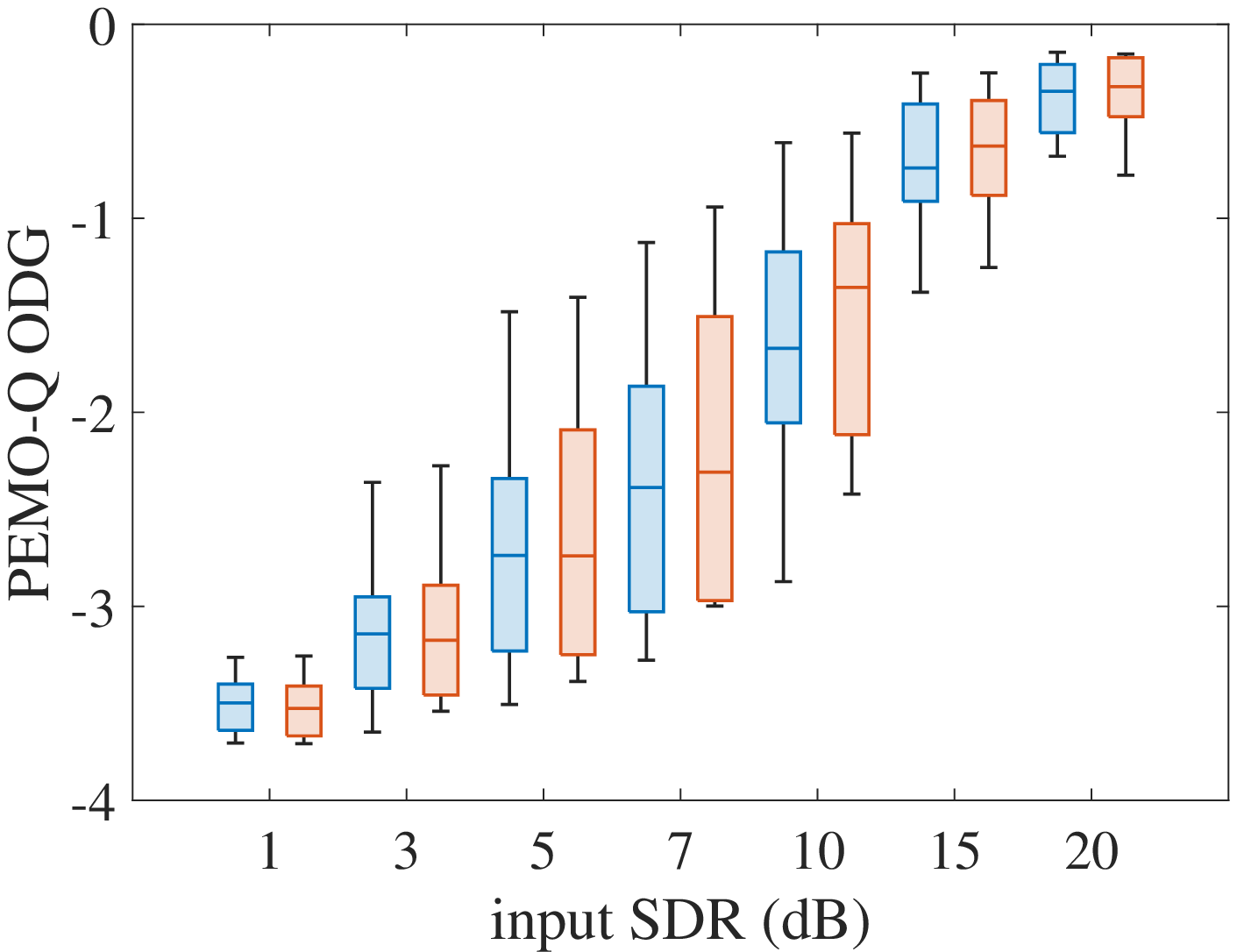}\label{subfig:PEMO_Q_PEW}}
\caption{Comparison of the synthesis and the analysis approaches to audio declipping for four shrinkage operators using PEMO-Q.}%
\label{fig:PEMO_Q_boxplots}%
\end{figure}

The PEMO-Q results shown in Fig.\,\ref{fig:PEMO_Q_boxplots} differ from the results obtained using the \dsdrc{} metric.
While the analysis variant in combination with the WGL shrinkage operator still remains marginally worse than its synthesis counterpart, 
for both the L and the PEW operators it slightly outperforms the synthesis variant in most of the test cases.
The LV algorithm using the EW shrinkage even marginally overcomes the FISTA algorithm utilizing PEW 
(which was also confirmed by informal listening tests), 
while being by about 6\,\% faster.
Nevertheless, note that the computer implementation used is general for all shrinkage types,
thus EW is run as a~special case of PEW, the size of the neighborhood $\mathcal{N}(z_{ft})$ being 1$\times$1.
Therefore, the computational complexity of EW could be further reduced by omitting the unnecessary steps related to neighborhood processing.

As mentioned earlier in this paper, 
the audio quality results of both the synthesis and analysis variants of social-sparsity declippers can be further 
improved via crossfaded replacement (CR) method
\cite{ZaviskaRajmicMokry2022:Declipping.crossfading} 
since the declipping methods produce results inconsistent in the reliable part.
This improvement tends to be slightly better in the analysis case.

\subsection{Coefficients weighting}

As mentioned in the introduction, motivated by the quality of results from \cite{ZaviskaRajmicSchimmel2019:Psychoacoustics.l1.declipping} 
obtained by the penalization of higher frequencies using quadratic (parabolic) weights,
we examined such a~weighting also in current experiments, i.e.\ when shrinkage operators other than the standard soft thresholding are used.
Note that the determination of these weights is computationally very cheap and therefore the weighting does not introduce any additional computational cost.

Formally, the vector of weights is computed as \mbox{$\w = \m \odot \m$}, 
where $\m = [1,\dots,\lfloor\frac{M}{2}\rfloor+1]$ and $M$ is the number of frequency bins of the STFT.
The weights are subsequently normalized to fit the range $[0, 1]$.

\begin{figure}[t]%
	\centering
	\vspace{-3mm}
	\hspace{3mm}
	\subfloat[][nonweighted]{
%
%
\definecolor{mycolor1}{rgb}{0.00000,0.44700,0.74100}%
\definecolor{mycolor2}{rgb}{0.85000,0.32500,0.09800}%
\definecolor{mycolor3}{rgb}{0.92900,0.69400,0.12500}%
\definecolor{mycolor4}{rgb}{0.49400,0.18400,0.55600}%
\begin{tikzpicture}[trim axis left, trim axis right, scale = 0.55]

\begin{axis}[%
width=2.7in,
height=1.9in,
at={(0.758in,0.481in)},
scale only axis,
xmin=0,
xmax=21,
xtick={ 1,  3,  5,  7, 10, 15, 20},
xticklabels={1,  3,  5,  7, 10, 15, 20},
xlabel style={font=\color{white!15!black}},
xlabel={input SDR (dB)},
ymin=0,
ymax=25,
ytick={0, 5, 10, 15, 20, 25},
yticklabels={0, 5, 10, 15, 20, 25},
ylabel style={font=\color{white!15!black}},
ylabel={$\Delta\text{SDR}_\text{c}\text{ (dB)}$},
ylabel shift = -1mm,
axis background/.style={fill=white},
xmajorgrids,
ymajorgrids,
axis line style={line width=0.25pt},
legend columns = 4,
legend style={at={(0.45,1.3)}, anchor=north west, legend cell align=left, align=left, draw=white!15!black, font=\small},
every axis plot/.append style={semithick}
]
\addplot [color=mycolor1, mark=x, mark options={solid, mycolor1}]
  table[row sep=crcr]{%
1	0.40896\\
3	1.2975\\
5	2.3972\\
7	3.8531\\
10	6.6008\\
15	9.4883\\
20	11.423\\
};
\addlegendentry{FISTA L}

\addplot [color=mycolor2, mark=x, mark options={solid, mycolor2}]
  table[row sep=crcr]{%
1	0.31556\\
3	1.2323\\
5	2.5245\\
7	4.4759\\
10	7.3237\\
15	9.9155\\
20	11.684\\
};
\addlegendentry{FISTA WGL}

\addplot [color=mycolor3, mark=x, mark options={solid, mycolor3}]
  table[row sep=crcr]{%
1	6.7553\\
3	8.6276\\
5	9.8611\\
7	10.814\\
10	12.162\\
15	13.884\\
20	14.9\\
};
\addlegendentry{FISTA EW}

\addplot [color=mycolor4, mark=x, mark options={solid, mycolor4}]
  table[row sep=crcr]{%
1	12.166\\
3	15.03\\
5	16.605\\
7	17.661\\
10	19.012\\
15	21.167\\
20	22.207\\
};
\addlegendentry{FISTA PEW}

\addplot [color=mycolor1, dashed, mark=o, mark options={solid, mycolor1}]
  table[row sep=crcr]{%
1	0.25534\\
3	1.0208\\
5	2.1106\\
7	3.6732\\
10	6.4924\\
15	9.3465\\
20	11.266\\
};
\addlegendentry{LV L}

\addplot [color=mycolor2, dashed, mark=o, mark options={solid, mycolor2}]
  table[row sep=crcr]{%
1	0.25968\\
3	1.0796\\
5	2.2491\\
7	3.9429\\
10	6.8009\\
15	9.454\\
20	11.19\\
};
\addlegendentry{LV WGL}

\addplot [color=mycolor3, dashed, mark=o, mark options={solid, mycolor3}]
  table[row sep=crcr]{%
1	8.3381\\
3	14.061\\
5	15.704\\
7	17.005\\
10	18.723\\
15	21.006\\
20	22.673\\
};
\addlegendentry{LV EW}

\addplot [color=mycolor4, dashed, mark=o, mark options={solid, mycolor4}]
  table[row sep=crcr]{%
1	8.4524\\
3	14.601\\
5	16.333\\
7	17.515\\
10	18.891\\
15	21.362\\
20	22.589\\
};
\addlegendentry{LV PEW}

\end{axis}
\end{tikzpicture}%
\label{subfig:dSDRclipped_plot_nonweighted}}\hfill
	\subfloat[][weighted]{
%
%
\definecolor{mycolor1}{rgb}{0.00000,0.44700,0.74100}%
\definecolor{mycolor2}{rgb}{0.85000,0.32500,0.09800}%
\definecolor{mycolor3}{rgb}{0.92900,0.69400,0.12500}%
\definecolor{mycolor4}{rgb}{0.49400,0.18400,0.55600}%
\begin{tikzpicture}[trim axis left, trim axis right, scale = 0.55]

\begin{axis}[%
width=2.7in,
height=1.9in,
at={(0.758in,0.481in)},
scale only axis,
xmin=0,
xmax=21,
xtick={ 1,  3,  5,  7, 10, 15, 20},
xticklabels={1,  3,  5,  7, 10, 15, 20},
xlabel style={font=\color{white!15!black}},
xlabel={input SDR (dB)},
ymin=0,
ymax=25,
ytick={0, 5, 10, 15, 20, 25},
yticklabels={0, 5, 10, 15, 20, 25},
ylabel style={font=\color{white!15!black}},
ylabel={$\Delta\text{SDR}_\text{c}\text{ (dB)}$},
ylabel shift = -1mm,
axis background/.style={fill=white},
xmajorgrids,
ymajorgrids,
axis line style={line width=0.25pt},
legend style={legend cell align=left, align=left, draw=white!15!black},
every axis plot/.append style={semithick}
]
\addplot [color=mycolor1, mark=x, mark options={solid, mycolor1}]
  table[row sep=crcr]{%
1	9.8544\\
3	12.518\\
5	14.26\\
7	15.45\\
10	16.834\\
15	18.924\\
20	20.144\\
};

\addplot [color=mycolor2, mark=x, mark options={solid, mycolor2}]
  table[row sep=crcr]{%
1	10.458\\
3	13.074\\
5	14.783\\
7	15.865\\
10	17.226\\
15	19.275\\
20	20.443\\
};

\addplot [color=mycolor3, mark=x, mark options={solid, mycolor3}]
  table[row sep=crcr]{%
1	7.6953\\
3	10.063\\
5	11.673\\
7	12.727\\
10	14.215\\
15	15.941\\
20	16.962\\
};

\addplot [color=mycolor4, mark=x, mark options={solid, mycolor4}]
  table[row sep=crcr]{%
1	5.366\\
3	7.3332\\
5	9.7424\\
7	11.316\\
10	14.09\\
15	16.317\\
20	18.062\\
};

\addplot [color=mycolor1, dashed, mark=o, mark options={solid, mycolor1}]
  table[row sep=crcr]{%
1	5.4798\\
3	8.8001\\
5	11.034\\
7	13.343\\
10	15.207\\
15	18.31\\
20	20.159\\
};

\addplot [color=mycolor2, dashed, mark=o, mark options={solid, mycolor2}]
  table[row sep=crcr]{%
1	5.6272\\
3	9.0981\\
5	11.426\\
7	13.752\\
10	15.323\\
15	18.128\\
20	19.87\\
};

\addplot [color=mycolor3, dashed, mark=o, mark options={solid, mycolor3}]
  table[row sep=crcr]{%
1	8.2752\\
3	10.639\\
5	12.543\\
7	14.077\\
10	15.981\\
15	18.899\\
20	20.578\\
};

\addplot [color=mycolor4, dashed, mark=o, mark options={solid, mycolor4}]
  table[row sep=crcr]{%
1	7.7571\\
3	9.7755\\
5	12.335\\
7	13.873\\
10	15.958\\
15	18.86\\
20	20.308\\
};

\end{axis}
\end{tikzpicture}%
\label{subfig:dSDRclipped_plot_weighted}} \\[-3mm]
	\hspace*{3mm}
	\subfloat[][nonweighted]{
%
%
\definecolor{mycolor1}{rgb}{0.00000,0.44700,0.74100}%
\definecolor{mycolor2}{rgb}{0.85000,0.32500,0.09800}%
\definecolor{mycolor3}{rgb}{0.92900,0.69400,0.12500}%
\definecolor{mycolor4}{rgb}{0.49400,0.18400,0.55600}%
\begin{tikzpicture}[trim axis left, trim axis right, scale = 0.55]

\begin{axis}[%
width=2.7in,
height=1.9in,
at={(0.758in,0.481in)},
scale only axis,
xmin=0,
xmax=21,
xtick={ 1,  3,  5,  7, 10, 15, 20},
xticklabels={1,  3,  5,  7, 10, 15, 20},
xlabel style={font=\color{white!15!black}},
xlabel={input SDR (dB)},
ymin=-4,
ymax=0,
ytick={ 0,  -1,  -2,  -3, -4},
yticklabels={0,  \textminus1,  \textminus2,  \textminus3, \textminus4},
ylabel style={font=\color{white!15!black}},
ylabel={PEMO-Q},
ylabel shift = -1mm,
axis background/.style={fill=white},
xmajorgrids,
ymajorgrids,
axis line style={line width=0.25pt},
legend style={legend cell align=left, align=left, draw=white!15!black},
every axis plot/.append style={semithick}
]
\addplot [color=mycolor1, mark=x, mark options={solid, mycolor1}]
  table[row sep=crcr]{%
1	-3.7486\\
3	-3.6864\\
5	-3.5941\\
7	-3.4342\\
10	-3.0169\\
15	-1.9691\\
20	-1.0522\\
};

\addplot [color=mycolor2, mark=x, mark options={solid, mycolor2}]
  table[row sep=crcr]{%
1	-3.759\\
3	-3.6739\\
5	-3.5572\\
7	-3.3386\\
10	-2.8369\\
15	-1.7134\\
20	-0.97077\\
};

\addplot [color=mycolor3, mark=x, mark options={solid, mycolor3}]
  table[row sep=crcr]{%
1	-3.7241\\
3	-3.6398\\
5	-3.5492\\
7	-3.3814\\
10	-3.0564\\
15	-2.065\\
20	-0.8883\\
};

\addplot [color=mycolor4, mark=x, mark options={solid, mycolor4}]
  table[row sep=crcr]{%
1	-3.4956\\
3	-3.1252\\
5	-2.6632\\
7	-2.3344\\
10	-1.6441\\
15	-0.75541\\
20	-0.38485\\
};

\addplot [color=mycolor1, dashed, mark=o, mark options={solid, mycolor1}]
  table[row sep=crcr]{%
1	-3.7564\\
3	-3.6852\\
5	-3.5702\\
7	-3.3868\\
10	-2.9028\\
15	-1.7817\\
20	-1.0333\\
};

\addplot [color=mycolor2, dashed, mark=o, mark options={solid, mycolor2}]
  table[row sep=crcr]{%
1	-3.7661\\
3	-3.6813\\
5	-3.57\\
7	-3.3796\\
10	-2.92\\
15	-1.7342\\
20	-1.0026\\
};

\addplot [color=mycolor3, dashed, mark=o, mark options={solid, mycolor3}]
  table[row sep=crcr]{%
1	-3.5652\\
3	-3.1136\\
5	-2.6633\\
7	-2.1951\\
10	-1.5357\\
15	-0.65266\\
20	-0.38577\\
};

\addplot [color=mycolor4, dashed, mark=o, mark options={solid, mycolor4}]
  table[row sep=crcr]{%
1	-3.5256\\
3	-3.0927\\
5	-2.5778\\
7	-2.1653\\
10	-1.4602\\
15	-0.66608\\
20	-0.36615\\
};

\end{axis}
\end{tikzpicture}%
\label{subfig:PEMO_Q_plot_nonweighted}}\hfill
	\subfloat[][weighted]{
%
%
\definecolor{mycolor1}{rgb}{0.00000,0.44700,0.74100}%
\definecolor{mycolor2}{rgb}{0.85000,0.32500,0.09800}%
\definecolor{mycolor3}{rgb}{0.92900,0.69400,0.12500}%
\definecolor{mycolor4}{rgb}{0.49400,0.18400,0.55600}%
\begin{tikzpicture}[trim axis left, trim axis right, scale = 0.55]

\begin{axis}[%
width=2.7in,
height=1.9in,
at={(0.758in,0.481in)},
scale only axis,
xmin=0,
xmax=21,
xtick={ 1,  3,  5,  7, 10, 15, 20},
xticklabels={1,  3,  5,  7, 10, 15, 20},
xlabel style={font=\color{white!15!black}},
xlabel={input SDR (dB)},
ymin=-4,
ymax=0,
ytick={ 0,  -1,  -2,  -3, -4},
yticklabels={0,  \textminus1,  \textminus2,  \textminus3, \textminus4},
ylabel style={font=\color{white!15!black}},
ylabel={PEMO-Q},
ylabel shift = -1mm,
axis background/.style={fill=white},
xmajorgrids,
ymajorgrids,
axis line style={line width=0.25pt},
legend style={legend cell align=left, align=left, draw=white!15!black},
every axis plot/.append style={semithick}
]
\addplot [color=mycolor1, mark=x, mark options={solid, mycolor1}]
  table[row sep=crcr]{%
1	-3.5851\\
3	-3.3067\\
5	-2.8296\\
7	-2.4437\\
10	-1.8869\\
15	-0.9754\\
20	-0.36762\\
};

\addplot [color=mycolor2, mark=x, mark options={solid, mycolor2}]
  table[row sep=crcr]{%
1	-3.5059\\
3	-3.1658\\
5	-2.6802\\
7	-2.2715\\
10	-1.7451\\
15	-0.92338\\
20	-0.3755\\
};

\addplot [color=mycolor3, mark=x, mark options={solid, mycolor3}]
  table[row sep=crcr]{%
1	-3.7017\\
3	-3.5867\\
5	-3.4423\\
7	-3.1984\\
10	-2.6607\\
15	-1.6171\\
20	-0.62809\\
};

\addplot [color=mycolor4, mark=x, mark options={solid, mycolor4}]
  table[row sep=crcr]{%
1	-3.7343\\
3	-3.6161\\
5	-3.4003\\
7	-3.0952\\
10	-2.4386\\
15	-1.4799\\
20	-0.57481\\
};

\addplot [color=mycolor1, dashed, mark=o, mark options={solid, mycolor1}]
  table[row sep=crcr]{%
1	-3.6154\\
3	-3.3469\\
5	-2.9255\\
7	-2.4219\\
10	-1.7764\\
15	-0.91882\\
20	-0.34828\\
};

\addplot [color=mycolor2, dashed, mark=o, mark options={solid, mycolor2}]
  table[row sep=crcr]{%
1	-3.6158\\
3	-3.3374\\
5	-2.9118\\
7	-2.4294\\
10	-1.8118\\
15	-0.96624\\
20	-0.39136\\
};

\addplot [color=mycolor3, dashed, mark=o, mark options={solid, mycolor3}]
  table[row sep=crcr]{%
1	-3.6157\\
3	-3.3614\\
5	-3.0251\\
7	-2.5091\\
10	-1.9305\\
15	-1.0179\\
20	-0.41126\\
};

\addplot [color=mycolor4, dashed, mark=o, mark options={solid, mycolor4}]
  table[row sep=crcr]{%
1	-3.6344\\
3	-3.4159\\
5	-3.1204\\
7	-2.6142\\
10	-2.0031\\
15	-1.0597\\
20	-0.4178\\
};

\end{axis}
\end{tikzpicture}%
\label{subfig:PEMO_Q_plot_weighted}}	
\caption{Comparison of plain (nonweighted) and weighted variants of the declipping algorithms, using \dsdrc{} and PEMO-Q metrics.}%
\vspace{-0.5em}
\label{fig:plots}%
\end{figure}
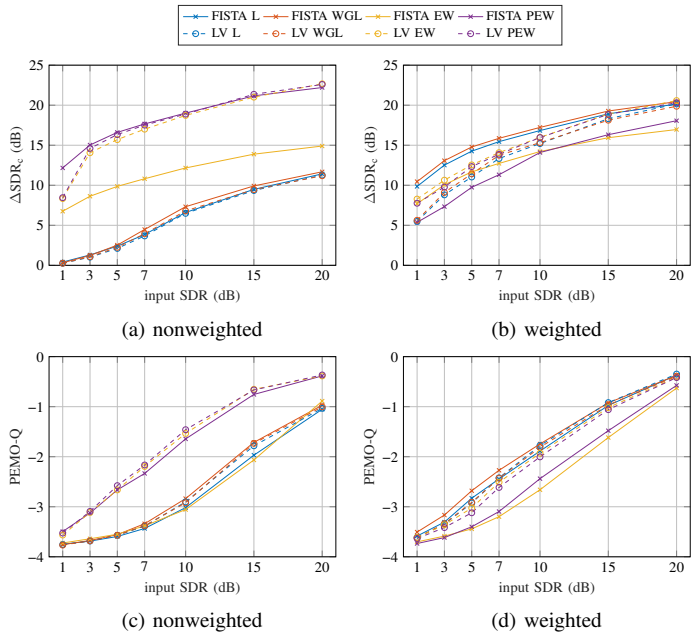

The results comparing the parabola-weighted and nonweighted variants of FISTA and LV declippers are illustrated in Fig.\,\ref{fig:plots}.
Herein, Fig.\,\ref{subfig:dSDRclipped_plot_nonweighted} shows the average \dsdrc{} values obtained from the nonweighted algorithms 
and Fig.\,\ref{subfig:dSDRclipped_plot_weighted} represents the parabola-weighted algorithms.
Different shrinkage operators are differentiated using colors, and algorithms are distinguished by line types (solid lines for FISTA, dashed lines for LV).

To summarize the results, it is possible to claim that parabola weighting improves the results for L and WGL.
However, such an improvement is more pronounced in the synthesis case.
Even though the synthesis variant with parabola-weighted WGL produces very good results, they are still worse by approximately 2\,dB compared with the results using PEW.
In the case of EW, the weighting slightly improves the results in the synthesis case but makes the results worse in the analysis case.
The analysis variant, nevertheless, remains better even when weighting is utilized.
For PEW, the results obtained when weighting is introduced are worse in both the synthesis and analysis cases.
Almost an identical conclusion can be drawn from the PEMO-Q results,
displayed in Figs.\,\ref{subfig:PEMO_Q_plot_nonweighted} and \ref{subfig:PEMO_Q_plot_weighted}.

\begin{figure}[h]
	\vspace{-1.5em}%
	\centering
	\subfloat[][magnitudes]{\hspace*{-1mm}\includegraphics[width=0.52\columnwidth]{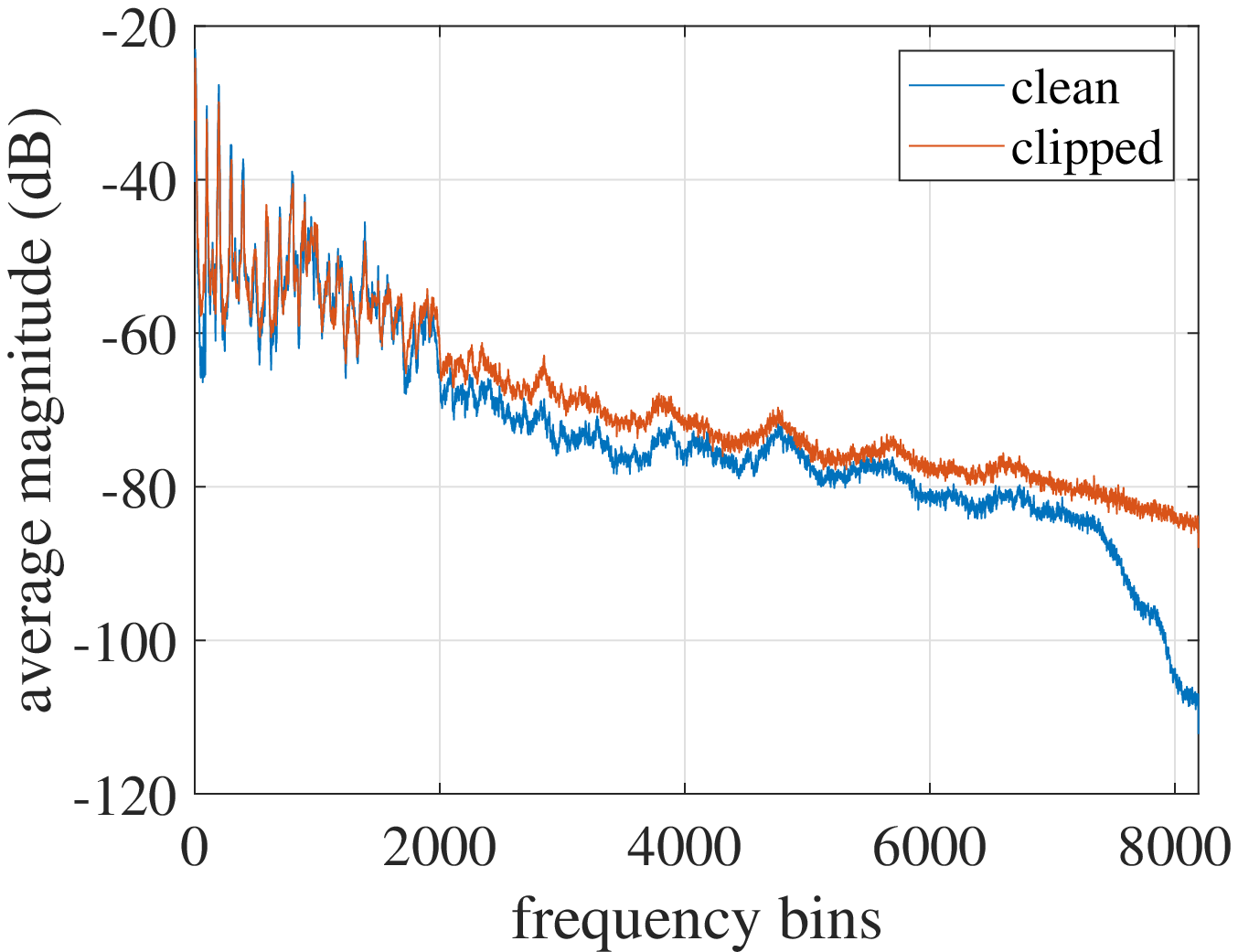}\label{subfig:magnitudes}}%
	\subfloat[][magnitude ratio]{\includegraphics[width=0.52\columnwidth]{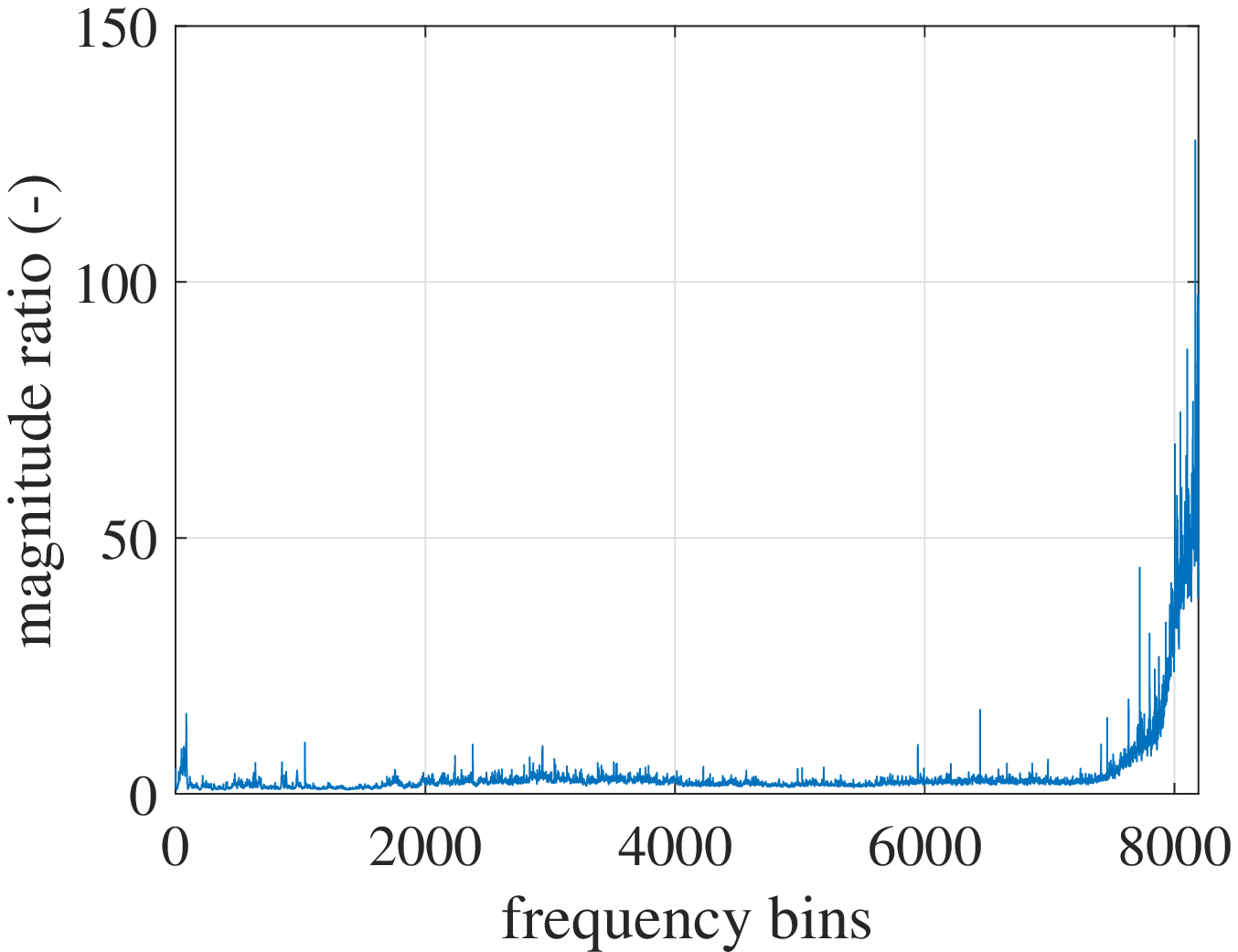}\label{subfig:magnitude_ratio}}
	\caption{Comparison of the magnitudes of STFT coefficients of clean and clipped audio signals.
	The magnitudes are obtained from the violin audio excerpt; the input SDR was 7\,dB.}
	\label{fig:magnitudes}
\end{figure}

\begin{figure}[t]%
	\vspace{-1.1mm}
	\centering
  \subfloat[][L]{\includegraphics[width=0.52\columnwidth]{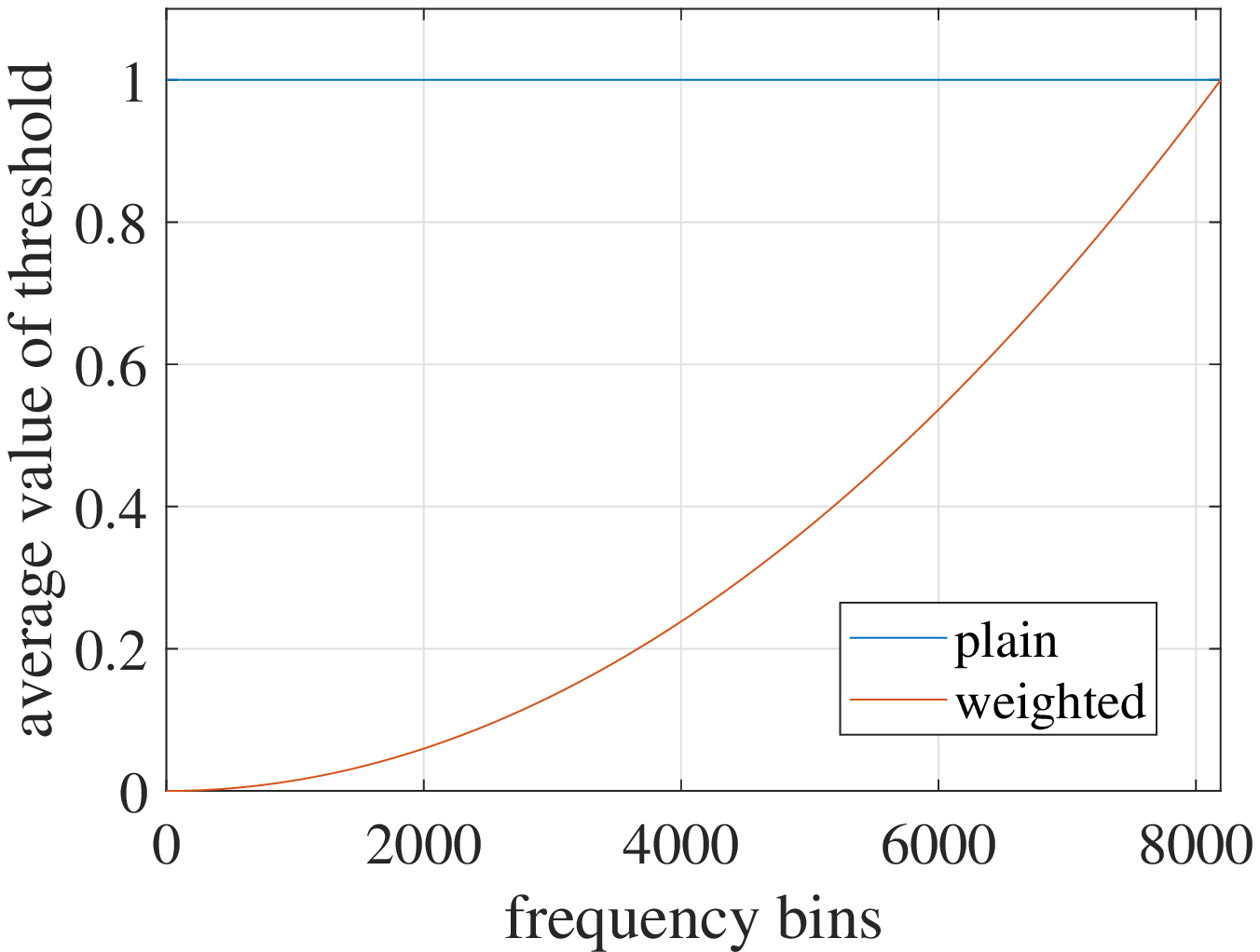}\label{subfig:threshold_L}}
	\subfloat[][WGL]{\includegraphics[width=0.52\columnwidth]{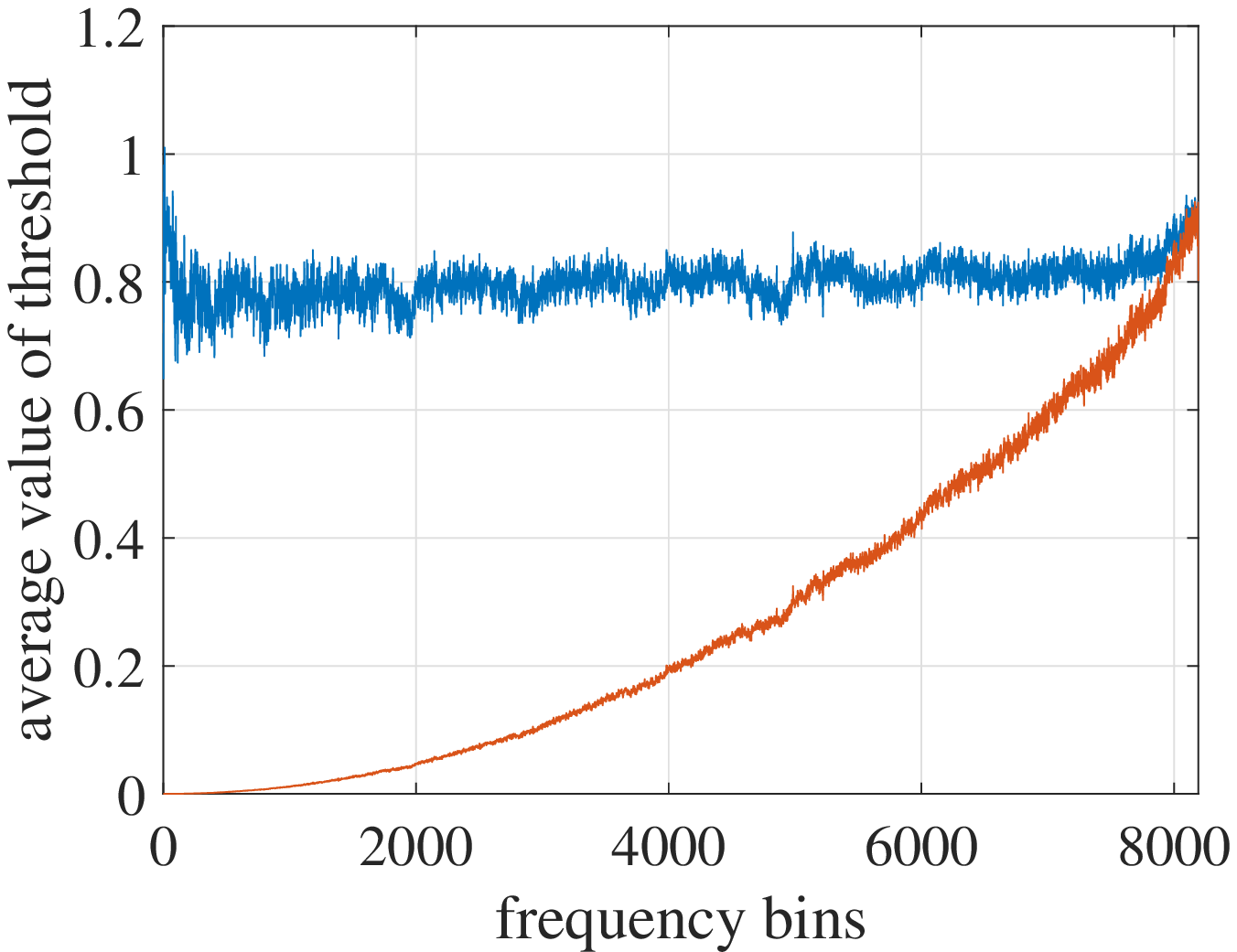}\label{subfig:threshold_WGL}}\\[-0.42cm]
	\subfloat[][EW]{\includegraphics[width=0.52\columnwidth]{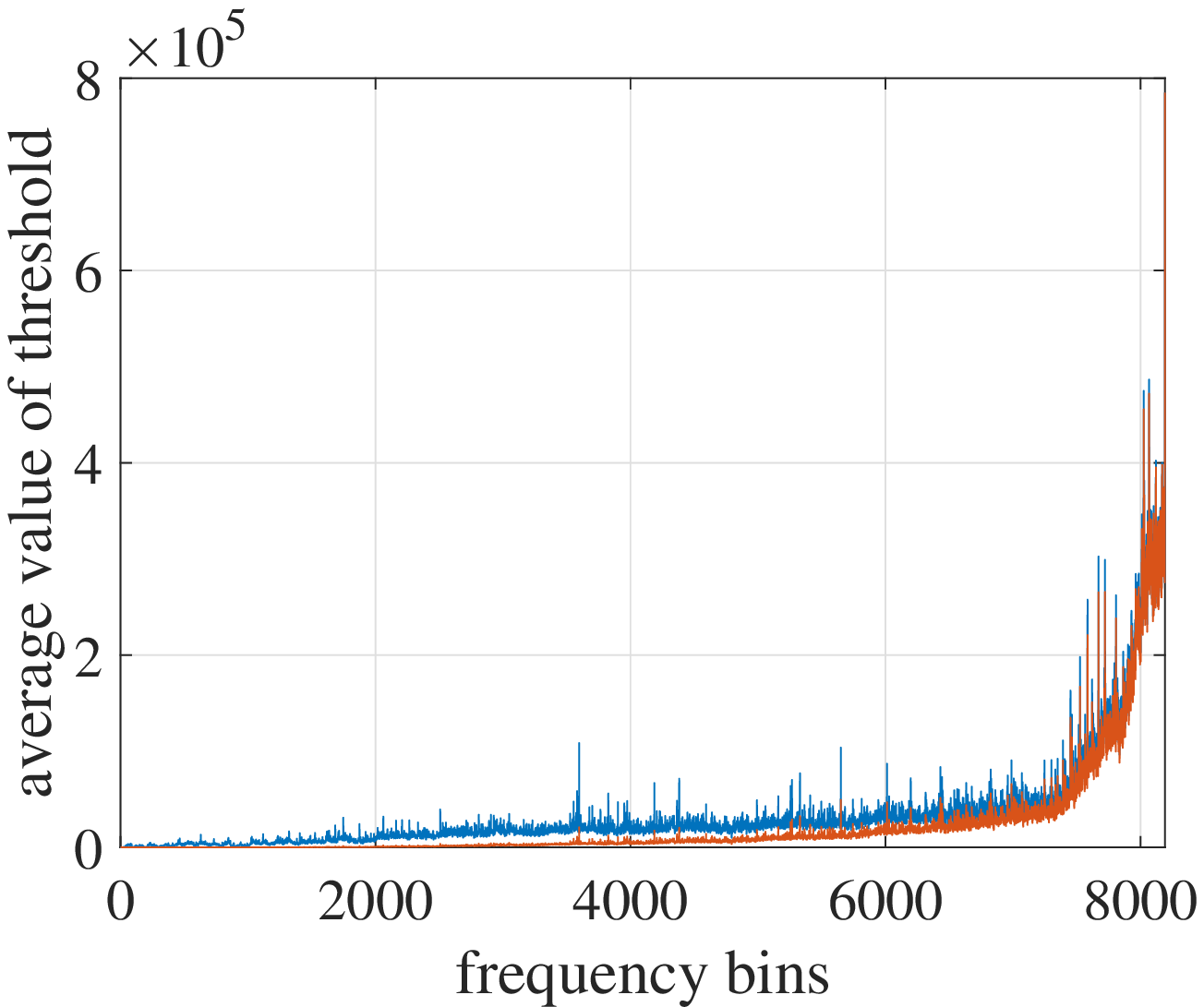}\label{subfig:threshold_EW}}
	\subfloat[][PEW]{\includegraphics[width=0.52\columnwidth]{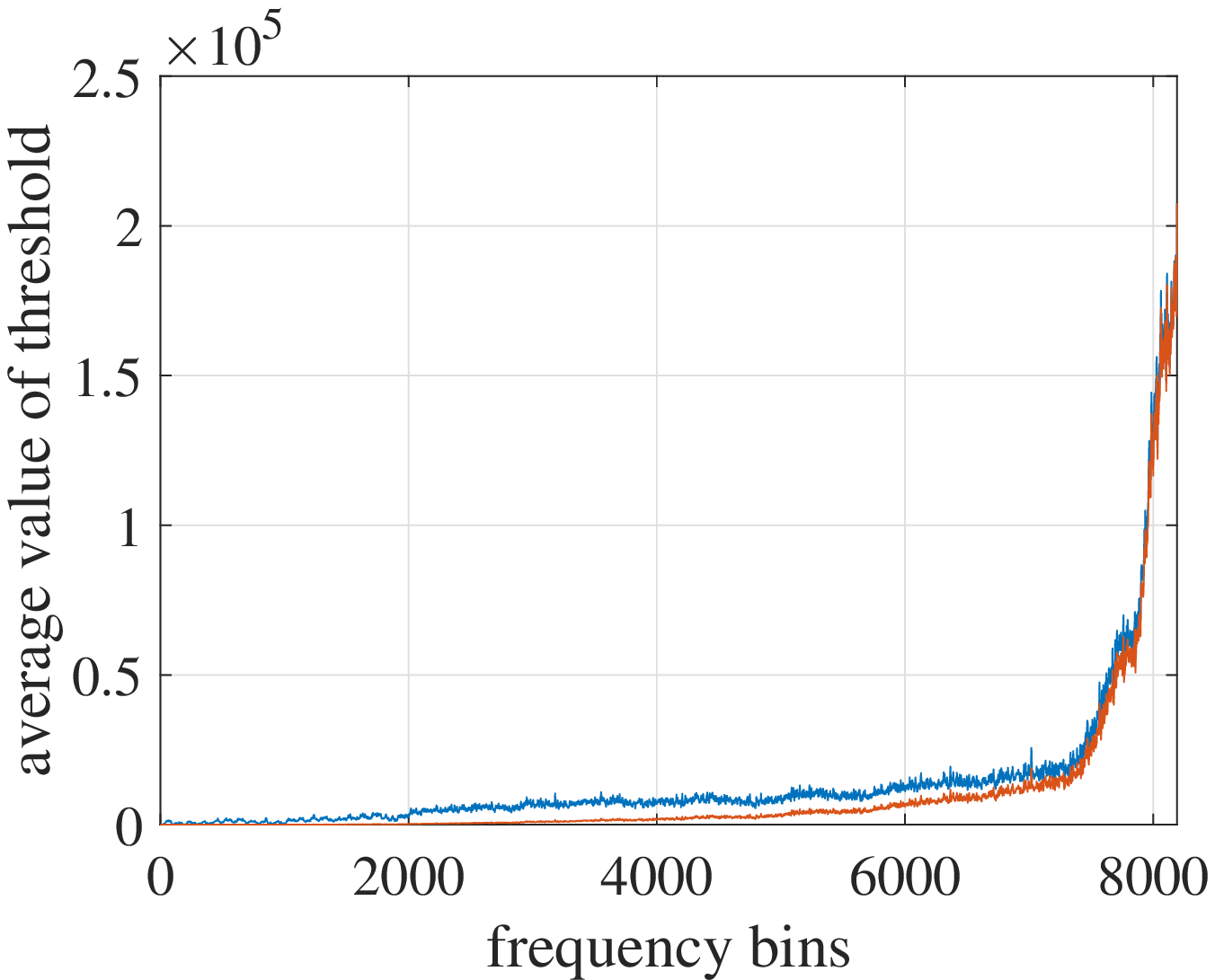}\label{subfig:threshold_PEW}}
	\vspace{0.5mm}
\caption{Average thresholds of shrinkage operators for each frequency bin.
Blue color marks thresholds of nonweighted variants and orange color represents thresholds exploiting the parabolic weights.}%
\vspace*{-1em}
\label{fig:thresholds}%
\end{figure}

The most probable reason for the above-mentioned behavior is that both EW and PEW are designed to promote the sparsity of the solution by suppressing smaller coefficients.
Since the energy of the time-frequency coefficients in audio signals generally decreases with frequency 
(see Fig.\,\ref{fig:magnitudes} for a~comparison of the magnitudes of the STFT coefficients), 
these shrinkages already suppress higher frequencies. 
Additional weighting to suppress higher frequencies even more may yield situations where the middle-band frequencies are not thresholded enough.
The average thresholds for each frequency bin for the violin audio excerpt are depicted in Fig.\,\ref{fig:thresholds}.

\section{Conclusion}
\vspace{-0.5mm}
We have proposed an analysis variant of the successful audio declipper.
For acquiring a~numerical solution, the Loris--Verhoeven algorithm has been adopted.
We have run experiments showing that the analysis variant brings improvement for some choices of the shrinkage operators.
The case of the Empirical Wiener shrinkage enjoyed significant gain in the reconstruction quality this way.
Most importantly, evidence shows that the analysis variant outperforms the synthesis counterpart uniformly (though slightly).

We furthermore generalized the social-sparsity declippers by including the weighting of the time-frequency coefficients.
The evaluation revealed that parabolic weights enhance the reconstruction quality in some setups (mostly L and WGL).
However, the best performing non-weighted choices were not beaten.
This effect has been explained.

The final preferences of algorithms that are expected to deliver top auditory quality within the examined family
(see Fig.\,\ref{subfig:PEMO_Q_plot_nonweighted}):
analysis PEW, analysis EW and synthesis PEW, all without weighting.
The analysis EW is moreover advantageous for its lower computational complexity.

The results presented are appended to the declipping survey website
\begin{center}
\url{https://rajmic.github.io/declipping2020/},
\end{center} 
through which the audio excerpts and MATLAB implementations are available.







%

{
\balance
\bibliographystyle{IEEEtran}
\inputencoding{cp1250}
\newcommand{\noopsort}[1]{} \newcommand{\printfirst}[2]{#1}
  \newcommand{\singleletter}[1]{#1} \newcommand{\switchargs}[2]{#2#1}

}


\end{document}